\documentclass{PoS}

\newcommand{\bea}{\begin{eqnarray}}
\newcommand{\eea}{\end{eqnarray}}

\newcommand{\bra}[1]{\left\langle #1 \right|}
\newcommand{\ket}[1]{\left| #1 \right\rangle}

\newcommand{\CMcal}{\mathcal}
\newcommand{\bear}{\begin{eqnarray}}
\newcommand{\eear}{\end{eqnarray}}
\newcommand{\nn}{\nonumber}
\newcommand{\beqn}{\begin{eqnarray}}
\newcommand{\eeqn}{\end{eqnarray}}
\newcommand{\be}{\begin{equation}}
\newcommand{\ee}{\end{equation}}

\newcommand{\ba}{\begin{array}{c}}
\newcommand{\bat}{\begin{array}{cc}}
\newcommand{\ea}{\end{array}}

\newcommand{\bi}{\begin{itemize}}
\newcommand{\ei}{\end{itemize}}

\newcommand{\Frac}[2]{\frac{\displaystyle #1}{\displaystyle #2}}

\newcommand{\cO}{{\cal O}}

\newcommand{\tQ}{\rm  Q}

\ShortTitle{
Anomalous \boldmath $AV^*V$ Green's function in soft-wall  AdS/QCD}

\abstract{

In this talk we study  the Green's function of two vector and one axial-vector currents
within  the soft-wall anti-de-Sitter (AdS)  model of Qunatum Chromodynamics (QCD),
with a quadratic dilaton and chiral symmetry broken through a field $X$
which gains a vacuum expectation  value.
We compare our predictions at high energies with the
Operator Product Expansion both in the massless quark limit and for $m_q\neq 0$.
The soft-wall model yields  a zero magnetic susceptibility $\chi=0$
and some problems are found in the case with $m_q\neq 0$.
We also discuss the relation proposed by Son and Yamamoto between the
$AV^*V$ and $VV-AA$ correlators, which is not obeyed at high energies
in   soft wall AdS/QCD.

}

\FullConference{
Sixth International Conference on Quarks and Nuclear Physics,  \\
April  16-20 2012, Palaiseau, France.
\\
\vspace*{0.5cm} This work was supported in part by  the Universidad CEU Cardenal Herrera
grants PRCEU-UCH15/10 and PRCEU-UCH35/11
and the MICINN-INFN fund AIC-D-2011-0818.  I would like to thank the organizers for the
kind scientific environment during the conference and all their help.
}

\title{\center{
Anomalous  $AV^*V$ Green's function\\ in soft-wall  AdS/QCD}}

\author{Juan Jos\'e Sanz-Cillero   \\
Istituto Nazionale di Fisica Nucleare, Sezione di Bari, Italy
E-mail: juan.sanzcillero@ba.infn.it}

\begin{document}

\section{Introduction:  $AV^*V$ Green's function }
\label{sec:intro}

The $AV^*V$ Green's function was recently studied
in the framework of soft-wall anti-de-Sitter (AdS)  theories~\cite{AVV-Colangelo}.
This analysis was motivated by a previous work by  Son and Yamamoto~\cite{Son:2010vc}
for holographic  theories where chiral symmetry is broken through
boundary condition~\cite{Son:2003+others}.   %%%,Sakai-Sugimoto,Hirn:2005nr}.
In Ref.~\cite{Son:2010vc},
the authors found an interesting relation between the $VV-AA$  correlator and
the Green's  function involving two  vector currents
$J_\mu=\bar q V \gamma_\mu q$
and
%%%the electromagnetic current
$J^{em}_\sigma=\bar q \tQ \gamma_\sigma    q$
and  an axial-vector current  $J^5_\nu=\bar q A \gamma_\nu \gamma_5 q$,
%%%with quark fields $q^i_f$ carrying a color $(i)$ and a flavour $(f)$ index, and
with $V$ and  $A$  diagonal matrices and the electric charge matrix $\tQ$:
%%% acting  on the flavour indices:
%
%We define the two-point correlation function of $J_\mu$ and $J^5_\nu$
%%%The QCD matrix element is defined  in a soft  external electromagnetic field,   with $k\to 0$:
\bear
T_{\mu \nu}(q,k)  &=&  i \,\, \int d^4x \, e^{i\,q \cdot x}
\bra{0} T[J_\mu(x)J_\nu^5(0)] \ket{\gamma(k,\epsilon)} %%%\,\,\, ,
\nn\\
&&
=\, - \, \Frac{i \, Q^2}{4 \pi^2} \, {\rm Tr}\left[\tQ V A\right] \,
P_\mu^{T\,\, \alpha}(q)
\, \left\{\,
 P_\nu^{T\,\,\beta}(q)\, w_T(Q^2) \,+  P_\nu^{L\,\,\beta}(q)  \,  w_L(Q^2) \,
\right\}\,   \, \tilde{f}_{\alpha\beta} \,\,,
\label{decomposition}
\label{twopoint}
\eear
with $k\to 0$ and  related to  the three-point   Green's function
$\bra{0} T[J_\mu(x)J_\nu^5(0)J^{em}_\sigma(y)] \ket{0}$.
%%%,
%%%with the  the electromagnetic current
%%%$J^{em}_\sigma=\bar q \tQ \gamma_\sigma    q$~\cite{VVA-Vainshtein}.
%%%
%%%vacuum correlation function
%%%\be
%%%  T_{\mu \nu \sigma} (q,k)=i^2  \,\, \int d^4x \, d^4y\, e^{i\,q \cdot x-i \, k \cdot y}
%%%  \bra{0} T[J_\mu(x)J_\nu^5(0)J^{em}_\sigma(y)] \ket{0}\,\,
%%%\label{threepoint}
%%%\ee
%%%\be
%%% T_{\mu \nu}(q,k)= e \,\epsilon^{ \sigma} \,T_{\mu \nu \sigma}(q,k) \,\,\, ,
%%%\label{threepoint-relation}
%%%\ee
%%%with $\epsilon^\sigma(k)$  the photon polarization vector
%%%and $e$ the electric charge unit.
%%%
%%%For soft photon momentum $k\to 0$ one can express  $T_{\mu \nu}(q,k)$
%%%keeping only linear terms  in $k$ and neglecting  quadratic and higher order powers of the
%%%photon momentum.
%%%In this kinematical condition, accounting for the conservation of the vector current $J_\mu$,
%%%The amplitude $ T_{\mu \nu}$ can be decomposed
%%%in terms of two structure functions $w_L(Q^2)$  and $w_T(Q^2)$:
%%%
%%%\be
%%%T_{\mu \nu}(q,k)=-{i \, \over 4 \pi^2} {\rm Tr}\left[Q V A\right] \left\{ w_T(q^2)(-q^2 {\tilde f}_{\mu \nu}+q_\mu q^\lambda {\tilde f}_{\lambda \nu} -q_\nu q^\lambda {\tilde f}_{\lambda \mu})+ w_L(q^2) q_\nu q^\lambda {\tilde f}_{\lambda \mu} \right\}\,\,, \label{decomposition}
%%%\ee
%%%
%%%\be
%%%T_{\mu \nu}(q,k)\, =\, - \, \Frac{i \, Q^2}{4 \pi^2} \, {\rm Tr}\left[\tQ V A\right] \,
%%%P_\mu^{T\,\, \alpha}(q)
%%%\, \left\{\,
%%% P_\nu^{T\,\,\beta}(q)\, w_T(Q^2) \,+  P_\nu^{L\,\,\beta}(q)  \,  w_L(Q^2) \,
%%%\right\}\,   \, \tilde{f}_{\alpha\beta} \,\,, \label{decomposition}
%%%\ee
We use the notation  $Q^2\equiv -q^2$,    ${\tilde f}_{\mu \nu}=
\displaystyle{1 \over 2} \epsilon_{\mu \nu \alpha \beta}  f^{\alpha \beta}$
and   %%%being the  dual field  of the photon field strength
$f^{\alpha \beta}=k^\alpha \epsilon^\beta -k^\beta \epsilon^\alpha$,
and  the transverse and longitudinal projectors,
respectively,
$P_{\mu\alpha}^{T}(q)=\eta_{\mu\alpha} -q_\mu q_\alpha /q^2$ and
$P_{\mu\alpha}^L=q_\mu q_\alpha /q^2$.

At short-distance  it is possible to use the Operator Product Expansion
(OPE) for $m_q=0$~\cite{AVV-Vainshtein,Knecht:2003xy}:
\begin{eqnarray}
w_L(Q^2)\, = \, \Frac{2 N_C}{Q^2}\, , \qquad
w_T(Q^2)  \,=\,  \frac{N_C}{Q^2}\,\,
+\,\, \frac{128 \pi^3 \alpha_s \,\chi\,\langle\bar{q}q\rangle^2}{9\, Q^6}\, \,\,\,
+\,\,\,\, {\CMcal O}\left(\frac{\Lambda^6}{Q^8}\right)\, .
\label{eq.AVV-OPE-meq0}
\end{eqnarray}
where the longitudinal component is completely fixed by the anomaly and does not receive
any correction~\cite{AVV-Vainshtein,Knecht:2003xy,Adler}
and $\chi$ is defined  by  the condensate
$\langle 0| \bar{q}\sigma^{\alpha\beta} q|\gamma\rangle =
i e  \chi \langle 0|  \bar{q} q|0\rangle f^{\alpha\beta}$.
%%%
%%%.
%%%The QCD dynamics is encoded in the perpendicular part, with the $D=4$ condensate
%%%determined  by   $\chi$,  defined  from the condensate
%%%$\langle 0| \bar{q}\sigma^{\alpha\beta} q|\gamma\rangle =
%%%i e  \chi \langle 0|  \bar{q} q|0\rangle f^{\alpha\beta}$.
%%%
%%%
%%%The magnetic susceptibility of the chiral condensate ($\chi$)
%%% arises  here after assuming factorization of the matrix element of  four-quark operators
%%%in a soft external electromagnetic field.
%%%
%%%In principle,  there might  be other ${\CMcal O}(1/Q^6)$ contributions in the OPE
%%%from  operators like $\widetilde{F}^{\alpha\beta} G^a_{\mu\nu}G_a^{\mu\nu}$,
%%%with $G^a_{\mu\nu}$ the gluon field strength;
%%%however, they appear at one-loop  with small coefficients,
%%%while the $1/Q^6$ term in~(\ref{eq.wT-OPE-mq0})
%%%comes from  tree-level diagrams.
%%%

If we allow $m_q\neq 0$,
%%%both longitudinal and perpendicular components get
the OPE yields corrections proportional to the quark mass at one loop~\cite{AVV-Vainshtein}:
\be
w_L(Q^2)-2 w_T(Q^2) = \cO\bigg(\Frac{\Lambda^4}{Q^6}\bigg)\, , \quad
w_T(Q^2) =  \Frac{N_C}{Q^2} \left[  1  +
\Frac{2 m_q^2}{Q^2}\, \ln{ \Frac{m_q^2}{Q^2} }
  -    \Frac{8 \pi^2 m_q \langle {\bar q}q \rangle \chi }{ N_C Q^2}
 +  \cO\left({\Lambda^4 \over Q^4} \right)\right] \, .
\label{eq.AVV-OPE-mdif0}
\ee
%%%Furthermore, at the two-loop level,
%%%$w_L-2 w_T$  is  $\cO(m_q^2/Q^4)$,    but suppressed by $\alpha_S$~\cite{Melnikov:2006qb}.

%%%the difference between $w_L$ and $2 w_L$
%%%already shows up at $\cO(m_q^2/Q^4)$, but suppressed by $\alpha_S$~\cite{Melnikov:2006qb}.

\vspace*{-0.5cm}
\section{The holographic setup in AdS/QCD}

We will consider a gauged   \ \ $U(n_f)_R\otimes U(n_f)_L$ chiral symmetry
and the AdS  line element  \ \
%%%.  We considered the five-dimensional AdS space with line element
$ ds^2=g_{MN}dx^M dx^N={R^2 \over z^2}(\eta_{\mu \nu} dx^\mu dx^\nu-dz^2)$,
%%%\label{metric}
%%%\ee
with the coordinate  \ \   indices   \ \   $M,N=0,1,2,3,5$,       \ \
$\eta_{\mu \nu}=$diag$(+1,-1,-1,-1)$,  $R$
 the AdS curvature radius (set to unity from now on)  and the 5D coordinate being in the range
 $0^+ \le z < + \infty$.
%%% .
%%%In the model,  the fifth coordinate $z$ runs in the range $\epsilon \le z < + \infty$,
%%%with $\epsilon \to 0^+$.
The 5D Yang-Mills  action describing the  fields  ${\cal A}_{L,R}^M$
dual to the left and right currents $J_{L,R}^\mu$,
as well as  the scalar-pseudoscalar  field $X$,  is given by
\be
 S_{YM}={1 \over k_{YM}} \int d^5x \sqrt{g}e^{-\Phi} Tr\left\{ |DX|^2
 %%%-m_5^2 |X|^2
 - \mathcal{V}(X)
 -{1 \over 4 g_5^2} (F_L^2+F_R^2) \right\} \,\,\, ,
 \label{action-0}
 \ee
 with the field strength tensors $F_{L,R}^{MN}  =F_{L,R}^{MNa}T^a$,
%%%$ =\partial^M {\cal A}^N_{L,R}
%%% -\partial^N {\cal A}^M_{L,R}-i \left[ {\cal A}^M_{L,R}, {\cal A}^N_{L,R} \right]$,
 $T^a$ the $U(n_f)$ group generators, the $X$ field potential $\mathcal{V}(X)$   and
 $g$  the determinant of the metric tensor $g_{MN}$.
 We take the quadratic dilaton background
 $\Phi(z)=(c z)^2$,  chosen in order to recover  linear Regge trajectories for vector resonances,
and $k_{YM}$ is a parameter included to provide canonical $4d$ dimensions
 for the fields.
%%% The 5d mass of the field $X$  is fixed to $m_5^2=-3$ according to the AdS/CFT correspondence
%%% dictionary.
The covariant derivative acting on $X$  is defined as
%%% \be
$D^M X=\partial^M X-i{\cal A}_L^MX+iX {\cal A}^M_R$.           %%%\,\,\,,
%%% \ee
%%%hence for $X=0$ the left and right sectors in (\ref{action-0}) are decoupled.
%%%
The  gauge fields ${\cal A}_{L,R}^M$ are usually combined into  a vector field
$V^M=\displaystyle{{\cal A}_L^M +{\cal A}_R^M \over 2}$
and an axial-vector field  $A^M=\displaystyle{{\cal A}_L^M -{\cal A}_R^M \over 2}$.
The study of the vector and scalar correlators  at high energies allows one
to fix the constants in the Yang-Mills action:
 $\displaystyle k_{YM}={16 \pi^2 \over N_c}$ and
 $\displaystyle g_5^2={3 \over 4}$~\cite{Son:2005+others}.   %%%,Colangelo:2008us}.

In this kind of
approaches~\cite{Son:2005+others},   %%%,Colangelo:2008us},
one  introduce a spinless  field $X$ which is  dual to the quark bifundamental operator
${\bar q}_R^\alpha q_L^\beta$.   This field gains the v.e.v.
$X =\Frac{v(y)}{2}\, e^{2 i\pi}$~\cite{Son:2005+others}.    %%%,Colangelo:2008us}.
Chiral symmetry becomes  broken  when $v(y)\neq 0$,   as the left and right sectors of the theory
get connected to each other.
Moreover, a phase-shift $\pi$ is induced for the v.e.v. in the bulk when
the parallel axial-vector source  is switched on:
$\pi$  gets coupled to $A^\parallel$ in the equations of motion (EoM).
Thus, for the bulk to boundary (B-to-b)  propagators one finds  the EoM,
within the  gauge $V_z=A_z=0$,
\bear
&&\partial_y \left({e^{-y^2} \over y}  \,\, \partial_y V_\perp \right)
-{\tilde Q}^2 {e^{-y^2} \over y}V_\perp=0\, ,
\qquad \quad
%%%\label{EOM-V} \\
%%%&&
\partial_y \left({e^{-y^2} \over y}  \,\, \partial_y A_\perp  \right)
-{\tilde Q}^2 {e^{-y^2} \over y}A_\perp -{g_5^2 v^2(y) e^{-y^2} \over y^3}A_\perp=0
\nn
%%%\label{EOM-A}
\\
&&\partial_y \left({e^{-y^2} \over y} \,\, \partial_y  { A_\parallel}   \right)
+{g_5^2 v^2(y) e^{-y^2} \over y^3}  ({  \pi}-{ A_\parallel } )=0\, , \qquad\qquad
%%%\label{EOM-phi-pi-1} \\
%%%&&
{\tilde Q}^2 (\partial_y  {  A_\parallel } )
+{g_5^2 v^2(y) \over y^2}\partial_y { \pi} =0
\label{EOM-phi-pi-2} \, ,
%%%
%%%\label{EOM-A} \\
%%%&&\partial_y \left({e^{-y^2} \over y} \,\, \partial_y  {\tilde \phi}^a  \right)
%%%+{g_5^2 v^2(y) e^{-y^2} \over y^3}  ({\tilde \pi}^a-{\tilde \phi}^a)=0
%%%\label{EOM-phi-pi-1} \\
%%%&&{\tilde Q}^2 (\partial_y  {\tilde \phi}^a)
%%%+{g_5^2 v^2(y) \over y^2}\partial_y {\tilde \pi}^a=0
%%%\label{EOM-phi-pi-2} \,\,\, ,
%%%
\eear
with  $y\equiv cz$ and
${\tilde Q}^2\equiv Q^2 /c^2$.
%%% $Q^2\equiv -q^2$  ($Q^2>0$ represent  the  Euclidean momentum).
In momentum space the 5D  fields
$\tilde\phi(q,y)=- i\frac{q^\mu}{q^2}\tilde{A}^\parallel_\mu(q,y)$
and $\tilde\pi(q,y)$ are respectively related to the B-to-b propagators
$A_\parallel(q,y)$  and  $\pi(q,y)$~\cite{AVV-Colangelo}.
%%%
%%%through  $ {\tilde \phi} (q,y)=
%%%-i{q^\mu \over q^2}A_\parallel(q,y) A^\parallel_{\mu 0} (q)$
%%%and $ \displaystyle {\tilde \pi} (q,y)=-i{q^\mu \over q^2}\pi(q,y) A^\parallel_{\mu 0} (q)$.
%%%

The vector EoM can be analyticaly solved~\cite{AVV-Colangelo},
%%%\be
%%%V_\perp(Q^2,y)\,=\,
%%%\Gamma\bigg(1+\Frac{Q^2}{4 c^2}\bigg)\,  U\bigg( \Frac{Q^2}{4 c^2},0,y^2\bigg)\, ,
%%%\ee
but for the remaining EoM one needs to specify the v.e.v.  $v(y)$. Its asymptotic
behaviour close to the UV brane ($y\to 0$ in our choice of coordinates) is related to the
explicit (quark mass $m_q$) and spontaneous chiral symmetry breaking
(quark condensate $\sigma\propto \langle \bar{q}q\rangle$ in massless QCD):
\bear
v(y)\,\,\,\stackrel{y\to 0}{=}\,\,\,  \Frac{m_q}{c}\,  y \, +\, \Frac{\sigma}{c^3} \, y^3\, \,\,
+\,\cO(y^4)\, .
\eear
where the first terms of its power expansion in $y$     determine
the   behaviour of $w_{T,L}$  at high-energies~\cite{AVV-Colangelo}.

%%%This  UV behaviour of $v(y)$ when $y\to 0$
%%%and the first terms of its power expansion in $y$     determines
%%%the asymptotic  behaviour of the Green's function at high-energies~\cite{AVV-Colangelo}.

The QCD chiral anomaly will be provided by the Chern-Simons action and,
more precisely, the $AV^*V$ amplitude studied here will be provided
by the piece~\cite{AVV-Colangelo}
\bear
S_{\rm CS}\bigg|_{AV^*V} &=& 3 \kappa_{CS}\, \epsilon_{ABCDE}\, \int d^5x\, \,  \,
\mbox{Tr}\bigg[ A^A \{F_{(V)}^{BC}  , F_{(V)}^{DE}\} \bigg]
\,\,=\,\,
48\kappa_{CS} \,d^{ab} \,\tilde{F}_{em}^{\mu\nu}\,
\int d^5x\,\, A_\nu^b\, \partial_z V_\mu^a\, ,
\eear
with the group factor $d^{ab}=$Tr$[\tQ\{ T^a,T^b\}]$. This yields the structure functions
\bear
w_{L\,\,(T)}  (Q^2) \,=\, -\, \Frac{2 N_C}{Q^2}\, \int_0^\infty dy\,\,
A_{\parallel\,\, (\perp)} (Q^2,y)\,  \partial_y V_\perp(Q^2,y)\, .
%%%\qquad
%%%w_T(Q^2) \,=\, -\, \Frac{2 N_C}{Q^2}\, \int_0^\infty dy\,\, A_\perp(Q^2,y)\,
%%%\partial_y V_\perp(Q^2,y)\, ,
\eear
The global normalization is fixed {\it a posteriori} through $\kappa_{CS}= -\frac{N_C}{96\pi^2}$
in the case with $m_q=0$.
%%%
%%%All that remains is to compute the B-to-b propagators
%%%$V_\perp$, $A_{\parallel}$ and $A_\perp$.

\section{$w_{T,L}$   results    for $m_q=0$ and $m_q\neq 0$ }

In the massless quark limit one can demonstrate that $A^\parallel(Q^2,y)=1$~\cite{AVV-Colangelo}.
The perpendicular B-to-b propagators can
be solved perturbatively in $1/\tilde{Q}^2$ in the form
$A^\perp(Q^2,y) =\sum_{n=0}^\infty A_n^\perp(t) (1/\tilde{Q}^2)^n$
and  $V^\perp(Q^2,y) =\sum_{n=0}^\infty V_n^\perp(t) (1/\tilde{Q}^2)^n$,
with $t\equiv y Q/c$.    This yields                %%%for the $AV^*V$ Green's function
the high-energy expansion
\bear
w_L(Q^2) \, = \, \Frac{2 N_C}{Q^2}\, ,  \qquad
%%%w_T(Q^2) \, = \, \Frac{N_C}{Q^2}\, \bigg[ \, 1\, -\, \Frac{2 g_5^2 \tau \sigma^2}{Q^6}
w_T(Q^2) \, = \, \Frac{N_C}{Q^2}\, \bigg[ \, 1\, -\, \Frac{3 \tau \sigma^2}{2 Q^6}
\,+\, \cO\bigg(\Frac{\Lambda^8}{Q^8}\bigg)\, \bigg]\, ,
\label{eq.AVV-AdS-meq0}
\eear
with $\tau\simeq 2.7$ defined by the integral of Bessel functions provided in
Ref.~\cite{AVV-Colangelo}.
The parallel B-to-b  propagator  $A_\parallel=1$  ensures the recovery of the
OPE prediction for$w_L$, which becomes fully determined by the
boundary conditions. Conversely, the QCD dynamics  is contained in $w_T$.
The comparison with the OPE~(\ref{eq.AVV-OPE-mdif0}) leads to a vanishing
prediction for the magnetic susceptibility $\chi=0$.
%%%
%%%Notice that $w_L$ was implicitly used to fix the Chern-Simons coupling, $\kappa_{CS}= - N_C/(96\pi^2)$.

In the case with $m_q\neq 0$,  all the  B-to-b propagators
can be solved perturbatively  in the way we did for Eq.~(\ref{eq.AVV-AdS-meq0}),
gaining  corrections proportional
to the quark mass and  leading to the amplitudes
%%%
%%%gain corrections proportional
%%%to the quark mass.  At high energies, we can solve perturbatively
%%%the B-to-b propagators  $A^\parallel,\, A^\perp ,\, V^\perp$
%%%in the way we did for Eq.~(\ref{eq.AVV-AdS-meq0}), leading to the amplitudes
%%%
\bear
& w_L(Q^2) \, = \, \Frac{2 N_C}{Q^2}
%%%\, \bigg[\, 1\, -\, (1-\pi(Q^2,0))\, \Frac{g_5^2 m_q^2}{2 Q^2}
\, \bigg[\, 1\, -\, (1-\pi(Q^2,0))\, \Frac{3 m_q^2}{8 Q^2}
\, +\, \cO\bigg(\Frac{m_q\Lambda^3}{Q^4}\bigg)\,\bigg]
\, ,    &
\nn\\
&
%%%w_T(Q^2) \, = \, \Frac{N_C}{Q^2}\, \bigg[ \, 1\, -\, \Frac{  g_5^2 m_q^2}{3 Q^4}
w_T(Q^2) \, = \, \Frac{N_C}{Q^2}\, \bigg[ \, 1\, -\, \Frac{    m_q^2}{4 Q^4}
\,+\, \cO\bigg(\Frac{m_q \Lambda^3}{Q^4}\bigg)
\,+\, \cO\bigg(\Frac{\Lambda^6}{Q^6}\bigg)\, \bigg]\, .     &
\eear
As   $A^\parallel$ and $\pi$ EoMs are coupled,  the perturbative solutions for
$Q^2\to\infty$   depend on the UV boundary condition $\pi(Q^2,0)$.
%%%,  which in principle is left unfixed.
%
The comparison of the NLO term proportional to $m_q$ with the
OPE~(\ref{eq.AVV-OPE-mdif0})    yields again a vanishing magnetic susceptibility $\chi=0$.
The   $m_q^2$ terms is more cumbersome since the recovery
of the finite OPE  log  $m_q^2 \ln{\frac{m_q^2}{Q^2}}$ in   $w_L(Q^2)$
requires a logarithmic dependence on $Q^2$ of the UV boundary  condition $\pi(Q^2,0)$.
The transverse component of the amplitude is even more problematic
as the holographic model generates  an  $m_q^2/Q^2$  term   without logs and
it is impossible to  recover the finite logarithms from the OPE without including
any further ingredient to the theory.

\section{Checking the Son-Yamamoto relation}

This work was motivated by the relation proposed by Son and Yamamoto
for $m_q=0$~\cite{Son:2010vc}
in the kind of model where chiral symmetry is broken through boundary
conditions~\cite{Son:2003+others}:
\be
w_T(Q^2) \, -\, \Frac{N_C}{Q^2}\,\,\,=\,\,\,  \Frac{N_C}{F_\pi^2}
\Pi_{VV-AA}(Q^2) \, .
\label{eq.SY-rel}
\ee
Actually,   although this kind of models fulfills this relation for any energy,  the left-had and
right-hand sides of~(\ref{eq.SY-rel}) do not obey the expected OPE short distance
behaviour~\cite{Son:2010vc}:
$ w_T(Q^2) \,-\,\Frac{N_C}{Q^2}\, =\, \cO(e^{-Q})$,
$\Pi_{VV-AA}(Q^2)\,=\, \cO(e^{-Q})$.

%%%\bear
%%%w_T(Q^2) \,-\,\Frac{N_C}{Q^2}\, =\, \cO(e^{-Q}) \, ,\qquad
%%%\Pi_{LR}(Q^2)\,=\, \cO(e^{-Q})\, .
%%%\eear

In the type of models   where chiral symmetry is broken through a scalar-pseudoscalar field $X$
that gains a v.e.v.~\cite{Son:2005+others},     %%% ,Colangelo:2008us},
%%%e.g.,   the soft-wall model  studied in Ref.~\cite{AVV-Colangelo},
one gets the  right $1/Q^6$ behaviour for the $VV-AA$ correlator
but the subleading corrections
in the $AV^*V$ Green's function do not start at the expected
orders~\cite{Son:2010vc,AVV-Colangelo}:
\bear
w_T(Q^2) \,-\,\Frac{N_C}{Q^2}\, =\,
%%%-\, \Frac{ 2 N_C g_5^2 \sigma^2\tau }{Q^8}
-\, \Frac{ 3 N_C \sigma^2\tau }{2 Q^8}
\,+\,\cO\bigg(\Frac{\Lambda^8}{Q^{10}}\bigg) \, ,\qquad
\Pi_{VV-AA}(Q^2)\,=\,-\,\Frac{N_C \sigma^2}{10\pi^2 Q^6}
\, +\,\cO\bigg(\Frac{\Lambda^8}{Q^8}\bigg)\, .
\eear
Hence,   Son-Yamamoto relation~(\ref{eq.SY-rel})   is not fulfilled in this kind of
models at high energies~\cite{AVV-Colangelo,Son:2010vc}.

It is worthy to mention an interesting result:
%%%that may explain why, nevertheless, the
%%%low-energy determinations  seem to be in relatively good agreement with data:
if we saturate the two Weinberg sum-rules for $w_T(Q^2)-N_C/Q^2$ stemming from
the OPE~\cite{AVV-Vainshtein,Knecht:2003xy}
through  the lightest multiplet of vector and axial-vector resonances one gets the
minimal hadronical approximation (MHA)~\cite{MHA},
\bear
w_T(Q^2)\bigg|_{\rm MHA} \, -\, \Frac{N_C}{Q^2}\, \,\,=\,\,\,
-\,\Frac{N_C M_V^2 M_A^2}{Q^2\, (M_V^2+Q^2) \, (M_A^2+Q^2)  }\, \,\,=\,\, \,
\Frac{N_C}{F^2}\, \Pi_{VV-AA}(Q^2)\bigg|_{\rm MHA}\,   ,
\eear
which fulfills the Son-Yamamoto relation~(\ref{eq.SY-rel}).
%%%By construction both $w_{T}$ and $\Pi_{LR}$  obey the high-energy behaviour prescribed
%%%by the OPE.
Although the MHA may lead to inaccurate  short-distance determinations
it provides a fair estimate of the low-energy constants~\cite{Masjuan:2007}.
This may explain the reasonable agreement for the low-energy
relation
$C_{22}^W \,\,\,=\,\,\, -\, \frac{N_C}{32\pi^2 F^2} \, L_{10}$
~\cite{Knecht:2011wh}.

%%%    which derives  from~(\ref{eq.SY-rel}),
.

\section{Conclusions}

We have studied the $AV^*V$ Green's function in the soft-wall~\cite{AVV-Colangelo}.
When  $m_q=0$ one has the B-to-b propagators
$\pi =A_\parallel =1$. This ensures the exact recovery of the longitudinal
structure amplitude  $w_L(Q^2)=2N_C/Q^2$   prescribed
by QCD~\cite{AVV-Vainshtein,Knecht:2003xy,Adler}.
%%%
%%%: it is a topological quantity determined
%%%by the boundary conditions, independent of the EoM of the 5D fields.
%%%
On the other hand, the transverse component
%%% reproduces the leading term but the subdominant
 corrections predicted in the soft-wall model start  at $\cO(1/Q^8)$,
producing a zero magnetic susceptibility $\chi$.    This hints the need for  further
ingredients in our holographic description like, e.g.,
the inclusion  of  a five-dimensional  field $B^{MN}$
dual to the tensor operator  $\bar{q}\sigma^{\alpha\beta} q$~\cite{tensor-Cata}.

%%%In particular,
%%%the 5D field dual to the tensor operator $\bar{q}\sigma^{\alpha\beta} q$ was not included
%%%and the transition $\gamma\to\bar{q}\sigma^{\alpha\beta} q$ is absent.
%%%Some works have already performed some studies on the impact of adding this tensor
%%%dual~\cite{tensor-Cata}.

The case  $m_q\neq 0$ brings further problems.  One needs to specify the
value of   $\pi(Q^2,y)$
%%%phase-shift B-to-b  propagator in the  UV boundary,
at $y\to 0$ and  the study of the subleading  terms  in the OPE
proportional to $m_q \sigma$ yields again
$\chi=0$.
Thus, the problem of the $m_q$ corrections needs further understanding
which might be obtained from the longitudinal part of the
$\Pi_{AA}(Q^2)$ correlator.

%%%are well understood in AdS/QCD and, in our opinion, the longitudinal part of the axial-vector
%%%correlator $\Pi_{AA}^\parallel(Q^2)$ should be a mandatory study in any computation
%%%with $m_q\neq 0$. We plan to study it in future works and expect it to provide a better
%%%understanding of these issues.

We have also tested the Son-Yamamoto relation
between the $AV^*V$ Green's function and the $VV-AA$ correlator~\cite{Son:2010vc}.
The hard and soft-wall models
%%% where chiral symmetry is broken through a scalar-pseudoscalar field $X$  which gains a v.e.v.
show  problems at high energies and the OPE is not well
recovered~\cite{AVV-Colangelo,Son:2010vc}.
However, the low-energy relation between even and odd-sector low-energy
constants  $C_{22}^W=- \frac{N_C}{ 32\pi^2 F^2} L_{10}$ seems
to be reasonably well satisfied~\cite{Knecht:2011wh}.

%%%The reason for this might be that the Son-Yamamoto and the sum-rules
%%%for the $AV^*V$ and $VV-AA$ amplitudes are well fulfilled  with the minimal hadronical
%%%approximation, where one only considers the lightest vector and axial-vector
%%%multiplets~\cite{MHA,Masjuan:2007}.

\end{document}